 \documentclass[doublespacing]{elsart}

\usepackage{cite}

 \usepackage{graphicx}

\usepackage{amssymb}
\begin{document}
\journal{Physics Letters A}
\begin{frontmatter}


\title{Effect of External Magnetic Field on Critical Current for the Onset of Virtual Cathode Oscillations in Relativistic Electron Beams}

\author[SSU]{Alexander Hramov},
\ead{aeh@nonlin.sgu.ru}
\author[SSU]{Alexey Koronovskii},
\author[SSU]{Mikhail Morozov},
\author[SSU]{Alexander Mushtakov}
\address[SSU]{Faculty of Nonlinear Processes, Saratov State %
University, Astrakhanskaya, 83, Saratov, 410012, Russia}

\begin{abstract}
In this Letter we research the space charge limiting current value
at which the oscillating virtual cathode is formed in the
relativistic electron beam as a function of the external magnetic
field guiding the beam electrons. It is shown that the space charge
limiting (critical) current decreases with growth of the external
magnetic field, and that there is an optimal induction value of the
magnetic field at which the critical current for the onset of
virtual cathode oscillations in the electron beam is minimum. For
the strong external magnetic field the space charge limiting current
corresponds to the analytical relation derived under the assumption
that the motion of the electron beam is one-dimensional [High Power
Microwave Sources. Artech House Microwave Library, 1987.
Chapter~13]. Such behavior is explained by the characteristic
features of the dynamics of electron space charge in the
longitudinal and radial directions in the drift space at the
different external magnetic fields.
\end{abstract}


\end{frontmatter}

\section{Introduction}
\label{Section:Introduction}

Beams of charged particles have great importance for understanding
physical properties of plasmas, as well as for their technological
applications. High power relativistic electron beams are used in
plasma heating, inertial fusion, high-power microwave generation,
and other. One of the most important directions of research in the
high-power vacuum and plasma electronics is the study of the virtual
cathode oscillators (vircators). The vircator is a microwave
generator whose operation is based on oscillations of the virtual
cathode (VC) in the electron beam with the overcritical current
\cite{Sullivan:1987_VCO_Book,
TrueAeh:2003_2004_microwave_electronics_1_2engl,
Dubinov:2002_ReviewEngl}. In recent years vircators have bean widely
studied experimentally \cite{Sze:1990_2VC, Alyokhin:1994,
Sze:1985_VC_without_Magnit, Sze:1986_VC, Hendricks:1990_2VC,
Davis:1988_reditron2, Kalinin:2005_PhysPlazma_ENG,
Dubinov:2002_ReviewEngl}, by computer simulation
\cite{Jiang:1996_VCO, Jiang:1995_VircatorKlystron,
Kwan:1988_reditron1, Chen:2004_VCO_Coaxial,
Lin:1990_VC_electrostatic, Anfinogentov:1998_VCO_ENGL,
Egorov:2006_VCO_2Dengl, Dubinov:2002_ReviewEngl} and analytical
analysis \cite{Bogdankevich:1971_SCL, Genoni:1980_SCL, Woo:1987_VCO,
Jiang1995_Mech_VC}. Microwave radiation in vircators is generated
when the current of a charged particle beam exceeds the critical
(so-called space charge limiting) current $I_{SCL}$
\cite{Sullivan:1987_VCO_Book,
TrueAeh:2003_2004_microwave_electronics_1_2engl}. Therefore, in
order to understand the physical processes occurring in beams with
the VC, it is important to determine the critical current $I_{SCL}$
for the formation of the oscillating VC in the electron beam.

The problem of the limiting current for a beam injected into a diode
gap was first treated in works of Child and Langmuir
\cite{child:1911, Langmuir:1923, Langmuir:1924}. This problem was
later considered by Pierce \cite{Pierce:1944}, who considered
one-dimensional dynamics of the electron beam in the planar diode
gap with a fixed neutralizing background (the so-called Pierce
diode) and stated that the increase of the beam current in Pierce
diode can lead to instability (which came to be called the Pierce
instability \cite{Godfrey:1987, Lawson:1989_PierceDiode,
Matsumoto:1996, Filatov:2006_PierceDiode_PLA}) with the resulting
formation of the VC \cite{Sullivan:1987_VCO_Book}. Further studies
presented that the development of the Pierce instability in the
diode gap is accompanied by the various nonlinear oscillatory
phenomena, such as chaotic spatiotemporal oscillations and the
formation of coherent structures (see, for example,
\cite{Godfrey:1987, Klinger:2001_PlasmaChaos,
Koronovskii:2002_PierceEnglish, hramov:013123}). Let us note, that
the studies by Child and Langmuir were later generalized numerically
to beams and drift spaces having a more complicated geometry
\cite{Luginsland:1996_PhysRevLett_VCO}.

Detailed analysis of the critical current values of relativistic
electron beams with different geometries was given firstly by
Bogdankevich and Rukhadze \cite{Bogdankevich:1971_SCL}. The work of
Genoni and Proctor \cite{Genoni:1980_SCL} produces later a more
accurate relation for space charge limiting current (see also
\cite{Sullivan:1987_VCO_Book}). These results were obtained under
the assumption that the motion of the electron beam is
one-dimensional (or similarly, that the guiding electrons external
magnetic field is infinitely strong). The theoretical condition for
the space charge limiting current assuming one-dimensional electron
motion is \cite{Sullivan:1987_VCO_Book,
TrueAeh:2003_2004_microwave_electronics_1_2engl,
Bogdankevich:1971_SCL, Genoni:1980_SCL}
\begin{equation}\label{I_0_B=INFTY}
I_{SCL}=\frac{mc^3}{e}\frac{\left(\gamma_0^{2/3}-1\right)^{3/2}}{1+2\ln
(R/r_b)},\qquad \gamma_0=\frac1{\sqrt{\left(1-(v_0/c)^2\right)}},
\end{equation}
where $R$ is the drift tube radius, $r_b$ is the radius of solid
beam, $\gamma_0$ is the value of relativistic factor at injection in
the drift space, $v_0$ is the electron velocity at injection in
drift space, $c$ is the light speed,  $e$ is the electron charge,
and $m$ is the electron rest mass.

Nevertheless, the problem of how the critical current for the onset
of the VC in the electron beam depends on the external axial
magnetic field has been not completely studied. It appears therefore
worthwhile to examine this issue in order to gain clearer and deeper
insight into the physical processes occurring in charged particle
beams with the VC. It should be noted that in
Refs.~\cite{Lindsay:2002_VCO, Chen:2004_VCO_Coaxial,
Egorov:2006_VCO_2Dengl} the complex nonlinear dynamics of the
electron beams with the VC in the finite external magnetic field was
studied, but in those papers the problem of the dependence of the
critical current for the onset of the VC oscillations on the
parameters of the system, was not analyzed.

In the present Letter we report the results of numerical studies
of how the critical current $I_{SCL}$ for the formation of the
unsteady oscillating VC in the weakly relativistic electron beam
depends on the external axial magnetic field.

The Letter is organized as follows. In
Section~\ref{Section:GeneralFormalism} the mathematical model of
studied electron-beam system is briefly discussed. In
Section~\ref{Section:CriticalCurrentDependence} the results of
investigations of how the external magnetic field affects the
critical current at which the VC oscillations arises in the
electron beam are presented. In Section 4 we discuss the physical
processes that accompany the formation of the VC in the electron
beam and govern the dependence $I_{SCL}(B)$ described in
Section~\ref{Section:CriticalCurrentDependence} for different
values of the external magnetic field. In conclusion, we summarize
the main results discussed in our Letter.

\section{General Formalism}
\label{Section:GeneralFormalism}

The simulation model will be briefly described in this section. The
drift space for the electron beam is the closed finite-length
cylindrical waveguide region of length $L$ and radius $R$ with grid
electrodes at both ends. An axially-symmetrical monoenergetic solid
electron beam with the current $I$ and radius $R_b$ is injected at
the velocity $v_0$ through the left (entrance) grid and than can be
is extracted through the right (exit) grid or escapes to the side
wall of the drift tube. An external uniform guiding electrons
magnetic field with induction $B$ is applied along the waveguide
axis.

We consider a time-dependent 2.5D model in which the dynamics of the
electron beam in the drift space is described by solving a
self-consistent set of Vlasov and Poisson equations
\cite{Birdsall:1985_PlasmaBook, Anderson:1999_EM_Simulations}. The
Vlasov kinetic equation for electron beam motion analysis is solved
numerically by the particle method \cite{Birdsall:1985_PlasmaBook,
Potter:1973_BookCompPhysics, Hockney:1981}. The potential
distribution in the drift waveguide can thus be easily obtained by
numerical solving the Poisson's equation in cylindrical geometry
\cite{Birdsall:1985_PlasmaBook}.

The model used in our study does not provide a correct analysis of
the problem of the critical current of electron beams with the
normalized velocity of $\beta_0 = v_0/c \gtrsim 0.5$, because it
does not take into account the self-magnetic field of the electron
beam and the associated beam pinching effects. Therefore we restrict
ourselves to studying the critical currents of weakly relativistic
electron beams.

The equations describing the dynamics of the weakly relativistic
electron beam are formulated in terms of the following
dimensionless variables of  the potential $\phi$, the space charge
field $E$, the electron density $\rho$, the electron velocity $v$,
the spatial coordinates $z$ and $r$, and the time $t$:
\begin{equation}\label{norm}
\begin{array}{ll}
  \varphi' = \left({v^2_0/\eta_0}\right)\varphi,\quad E'=\left({v^2_0/L\eta_0}\right)E,
  \quad B'=\left({v_0/L\eta_0}\right)B,\quad \rho' =\rho_0\rho,  \\
  v'=v_0 v,\quad z'=L z,\quad r'=L r,\quad t'=\left({L/v_0}\right)t,  \\
\end{array}
\end{equation}
Here the prime means the dimensionless values, $\eta_0$ is the
specific charge of electron, $v_0$ and $\rho_0$ are the velocity and
space charge density of the electron beam at the entrance to the
system, respectively, and $L$ is the length of the interaction
space.

As mentioned above, unsteady processes in the electron beam injected
into the drift space were simulated numerically by the particle
method. In cylindrical geometry particles have the form of a charged
ring. For each of those charged particles the equations of motion
were solved with allowance for the weakly relativistic velocity of
the beam electrons. In terms of dimensionless values (\ref{norm}),
the equations of motion in cylindrical coordinates are written as
\begin{equation}
 \label{vlasov1}
\frac{d^2r_i}{dt^2}-r_i\left(\frac{d\theta_i}{dt}\right)^2=\eta(z_{i},\theta_{i},r_{i})\left(E_r+r_iB\frac{d\theta_i}{dt}\right),
\end{equation}
\begin{equation}
 \label{vlasov2}
r_i\frac{d^2\theta_i}{dt^2}+2\frac{d\theta_i}{dt}\frac{dr_i}{dt}=-\eta(z_{i},\theta_{i},r_{i})B\frac{dr_i}{dt},
\end{equation}
\begin{equation}
 \label{vlasov3}
\frac{d^2z_i}{dt^2}=\eta(z_{i},\theta_{i},r_{i})E_z, \qquad
i=1,\dots N_0,
\end{equation}
where
\begin{equation}
\eta(z_{i},\theta_{i},r_{i})=\left(1-\frac{\beta_0^{2}}{2}\left[\left(\frac{d
r_{i}}{d t}\right)^{2}+\left(r_{i}\frac{d\theta_{i}}{d
t}\right)^{2}+\left(\frac{d z_{i}}{d t}\right)^{2}\right] \right).
\end{equation}
Here, $z_i$, $r_i$, and $\theta_i$ are the longitudinal, radial, and
azimuthal coordinates of the charged particles, respectively, $E_z$
and $E_r$ are the longitudinal and radial electric field components,
respectively, $B = B_z$ is the longitudinal component of the
external magnetic field, whose radial component is assumed to be
zero $B_r = 0$, $\beta_0 = v_0/c$, where $v_0$ is the electron beam
velocity at the entrance to the system and $c$ is the light speed.
The fields do not have azimuthal components because the system is
axially-symmetric. The subscript $i$ denotes the number of particles
and $N_0$ is the full number of particles used to model the charged
particles beam.

The potential distribution in the interaction space was calculated
self-consistently from Poisson's equation
\begin{equation}
 \label{puasson}
\frac{1}{r}\frac{d\varphi}{dr}+\frac{d^2\varphi}{dr^2}+\frac{d\varphi^2}{dz^2}=\alpha^2\rho,
\end{equation}
where
\begin{equation}
 \label{ALPHA}
\alpha=L\left(\frac{|\rho_0|}{\varphi_0\varepsilon_0}\right)^{1/2}=\sqrt{2}\,\omega_pL/v_0
\end{equation}
is the dimensionless control parameter which depends on the beam
current as $\alpha \sim \sqrt{I}$ and is proportional to the
length of the drift tube space as $\alpha \sim L$. Poisson's
equation (\ref{puasson}) was solved with the boundary conditions
\begin{equation}
 \label{side1}
\varphi (z=0,r)=0, \qquad \varphi(z=1,r)=0, \qquad
\varphi(z,r=R)=0,
\end{equation}
\begin{equation}
 \label{side2}
\left.\displaystyle\frac{d\varphi}{dr}\right|_{r=0}=0.
\end{equation}
Condition (\ref{side1}) implies that drift region bounded by the
conducting cylindrical surface of the radius $R$ held at a zero
potential. Condition (\ref{side2}) means that the radial electric
field at the symmetry axis $r = 0$ is zero due to the axial
symmetry of the interaction space.

For each particle, equations of motion
(\ref{vlasov1})--(\ref{vlasov3}) were integrated numerically by the
second-order leap-frog method \cite{Birdsall:1985_PlasmaBook}. At
each time step Poisson's equation (\ref{puasson}) was solved on the
two-dimensional mesh in cylindrical coordinates. In order to reduce
the mesh noise, the space charge density on the mesh was calculated
using a bi-linear weighing procedure for particles (the
particle-in-cell (PIC) method) \cite{Birdsall:1985_PlasmaBook}.

\section{Critical current dependence on the
external magnetic field} \label{Section:CriticalCurrentDependence}

Let us consider the results of numerical study of how the external
magnetic field affects the space charge limiting current $I_{SCL}$
at which the VC oscillations arises. The value of the critical
current  is determined by the calculation of the trajectory of the
charged particles in the phase space $(\bf{r}, \bf{v})$, where $\bf
r$ is a radius-vector and $\bf v$ is a velocity of the charged
particle. The moment when virtual cathode appears corresponds to
reflection of a part of the electron beam backwards to the injection
plane $z=0$. In this case, appearance of the charged particles with
negative longitudinal components of velocity ($v_z<0$) takes place.
Onset of the oscillations of the virtual cathode is investigated by
the analysis of the dynamics of the space charge electric field
(both longitudinal $E_z$, and radial $E_r$ components) in the drift
space. The moment of appearance of charged particles reflections in
the beam (virtual cathode formation) corresponds to the moment of
onset of the electric field oscillations in the system. Let us note
that if an injected current exceeds the critical current value
insignificantly then the duration of the time interval required for
virtual cathode formation and the reflected particle occurrence
grows considerably (see \cite{Alyokhin:1994}). Therefore we took
into account this fact in the numerical simulation by means of
choice of the duration of the numerical simulation large enough for
the beam current being close to the value of the critical current.
So, in this case the duration of the simulation exceeded $200T_p$,
where $T_p = 2\pi/\omega_p$ is the typical time scale,  $\omega_p$
is the plasma frequency of the electron beam. This duration
corresponds approximately 100 periods of the virtual cathode
oscillations.

\begin{figure}[b]
\centerline{\includegraphics*[scale=0.4]{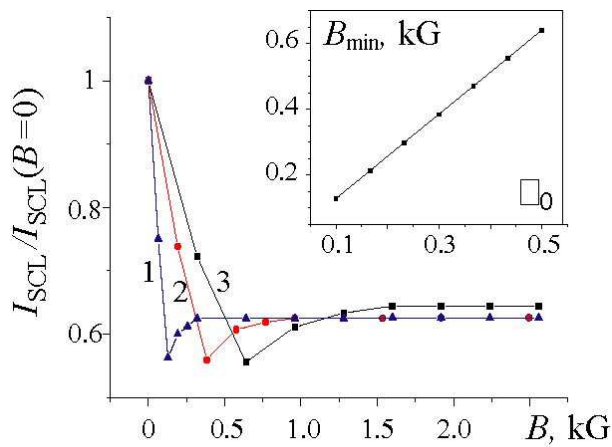}}
\caption{Normalized critical (space charge limiting) current
$I_{SCL}/I_{SCL}(B=0)$ vs external magnetic field $B$ for $\sigma =
0.5$ and for the different electron beam velocities: ({\sl1})
$\beta_0 = 0.1$, ({\sl2}) 0.3, and ({\sl3}) 0.5. In the frame the
optimum external magnetic field $B_{\min}$ vs normalized electron
velocity $\beta_0$ is shown. \label{fgr1}}
\end{figure}

\begin{figure}[t]
\centerline{\includegraphics*[scale=0.4]{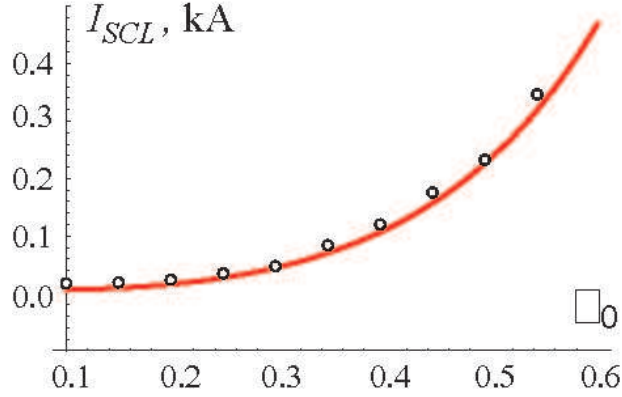}}
\caption{{Critical (space charge limiting) current $I_{SCL}$ [kA] vs
normalized electron beam velocity $\beta_0$ for $\sigma = 0.5$ and
strong magnetic field $B=2.5$\,kG (points $\circ$). The solid line
corresponds to the analytical relation (\ref{I_0_B=INFTY}) of the
space charge limiting current. \label{fgr1_2ADD}}}
\end{figure}

We begin by analyzing the case in which the ratio of the initial
beam radius $r_b$ at the entrance to the interaction space to the
radius $R$ of the interaction space itself $\sigma = r_b/R$ is
fixed. We call $\sigma$ the filling parameter, i.e. the parameter
that characterizes the extent to which the drift space is filled
with the electron beam. If it is not stipulated especially, the
ratio of the drift tube radius to its length is assumed to be
constant and is equal to $R/L = 0.25$. Fig.~\ref{fgr1} shows how the
normalized critical beam current $I_{SCL}$ depends on the external
magnetic field $B$ for $\sigma = 0.5$. The critical beam current is
normalized to its magnitude $I_{SCL}$ at a zero external magnetic
field ($B = 0$). In Fig.~\ref{fgr1} different curves correspond to
different initial velocities $\beta_0 = v_0/c$ of the electron beam
injected into the drift tube. We can see that the dependence
$I_{SCL}(B)$ behaves differently in two characteristic ranges of the
magnetic field $B$. At weak magnetic fields the critical beam
current decreases monotonically as the external magnetic field $B$
increases. At strong external magnetic fields the critical current
of the electron beam increases monotonically with magnetic
induction. Consequently, there is a certain magnetic field
$B_{\min}$ at which the critical beam current is minimum. This
result is quite important because it provides the way to determine
the optimum magnetic field for the onset of microwave generation in
the virtual cathode oscillator at the minimum current of
relativistic electron beam. Note, that at strong magnetic fields the
critical beam current increases and asymptotically approaches to the
analytical relation (\ref{I_0_B=INFTY}) of space charge limiting
current which was obtained for an infinitely strong external
magnetic field (and, as consequence, one-dimensional motion of the
electron flow) \cite{Sullivan:1987_VCO_Book}. {For verification of
the results of our simulation in Fig.~\ref{fgr1_2ADD} the analytical
(Eq.~(\ref{I_0_B=INFTY})) and numerical dependencies of the critical
current $I_{SCL}$ on the normalized electron beam velocity $\beta_0$
for strong magnetic field $B=2.5$\,kG are shown. Comparison of the
presented dependencies shows good agreement of the known analytical
results and our numerical calculations.}

\begin{figure}[t]
\centerline{\includegraphics*[scale=0.7]{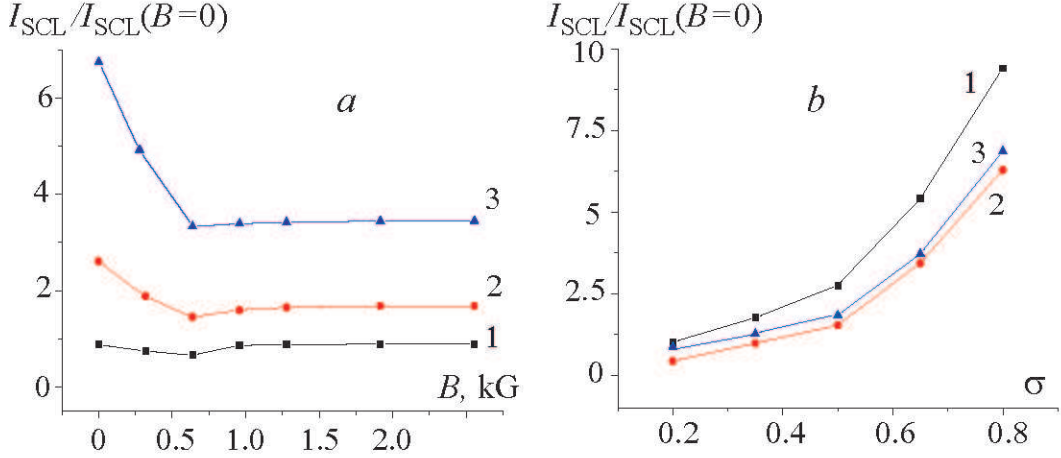}} \caption{(a)
Normalized critical electron beam current $I_{SCL}/I_{SCL}(B = 0)$
vs external magnetic field for different values of the filling
parameter $\sigma$ (the extent to which the drift tube is filled
with the beam). Curve~{\sl1} corresponds to $\sigma = 0.2$, {\sl2}
-- 0.5, and {\sl3} -- 0.9. (b) Normalized critical current vs.
filling parameter $\sigma$ for different inductions of the external
magnetic field: ({\sl1}) $B = 0$, ({\sl2}) $B=B_{\min}=0.6$\,kG, and
({\sl3}) $B = 2.5$\,kG. The electron beam velocity at the entrance
to the drift tube is such that $\beta_0 = 0.5$ \label{fgr2}}
\end{figure}

The dependencies shown in Fig.~\ref{fgr1} were calculated for
different values of the electron beam velocity $\beta_0$ at the
entrance to the drift tube. We can see from Fig.~\ref{fgr1}, that
for the non-relativistic and relativistic electron beams under
consideration here the dependence of the critical beam current
$I_{SCL}$ on the external magnetic field $B$ is qualitatively the
same for different beam velocities. For any velocity of the injected
beam the critical current decreases with increasing $B$ in the range
of the weak external magnetic fields, and, otherwise, it increases
monotonically in the range of the strong magnetic fields. However,
the optimal magnetic field $B_{\min}$ depends on the electron beam
velocity $\beta_0$. The dependence of $B_{\min}$ on $\beta_0$ is
shown in the frame in Fig.~\ref{fgr1} from which we can see that the
optimum external magnetic field increases linearly with the velocity
$\beta_0$ of the injected electron beam.

\begin{figure}[t]
\centerline{\includegraphics*[scale=0.7]{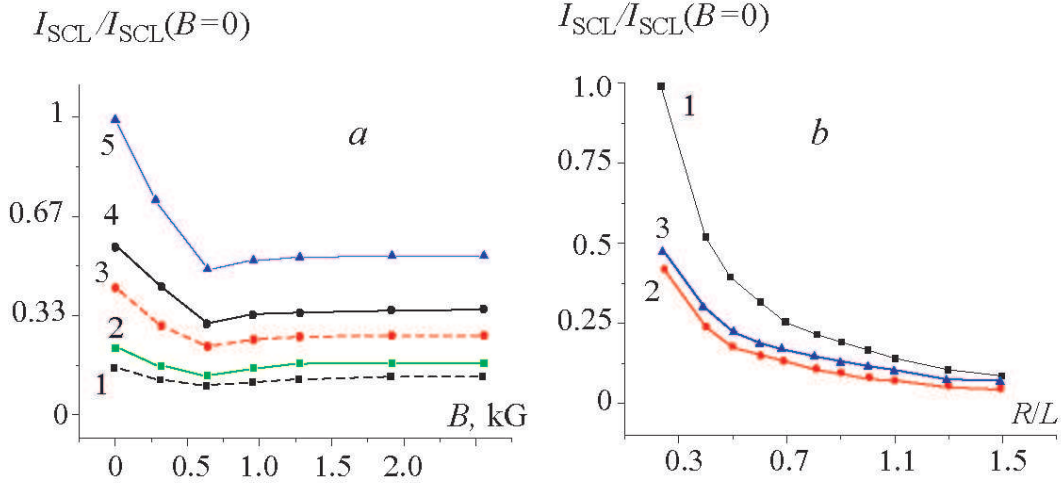}}
\caption{{(a) Normalized critical electron beam current
$I_{SCL}/I_{SCL}(B = 0)$ vs external magnetic field for different
values of the ratio of the drift tube radius to its length for fixed
beam radius $r_b/L=0.25$. Curve~{\sl1} corresponds to $R/L = 1.3$,
{\sl2} -- 1.0, {\sl3} -- 0.5, {\sl4} -- 0.4, and {\sl5} -- 0.3. (b)
Normalized critical current vs. the ratio $R/L$ for different
inductions of the external magnetic field: ({\sl1}) $B = 0$,
({\sl2}) $B=B_{\min}=0.6$\,kG, and ({\sl3}) $B = 2.5$\,kG. The
electron beam velocity at the entrance to the drift tube is such
that $\beta_0 = 0.5$ \label{fgr2_2ADD}}}
\end{figure}

Let us now consider the dependence of the critical current for the
formation of the VC oscillation on the external magnetic field for
different values of the filling parameter $\sigma$ {and the ratio
$R/L$ of the drift tube radius to its length}.

Fig.~\ref{fgr2}(a) presents the corresponding dependencies
$I_{SCL}(B)$ for different $\sigma$ values and for the fixed
electron beam velocity $\beta_0 = 0.5$ at the entrance to the
drift tube. We can see, that the current required for the
formation of the oscillating VC increases with the filling
parameter.  In Fig.~\ref{fgr2}(b) the critical beam current
dependencies on the filling parameter $\sigma$ for different
values of the external magnetic field are shown. It can be seen
that the critical beam current $I_{SCL}$ increases monotonically
with $\sigma$, and that for different external magnetic fields the
dependence $I_{SCL}(\sigma)$ remains essentially the same.

{Fig.~\ref{fgr2_2ADD}(a) shows the dependencies $I_{SCL}(B)$ for
different values the ratio $R/L$ of the drift tube radius to its
length and for the fixed electron beam velocity $\beta_0 = 0.5$. In
this case the current required for the formation of the oscillating
VC decreases with the growth of the $R/L$ value. In
Fig.~\ref{fgr2}(b) the critical current dependencies on the geometry
parameter $\sigma$ for different values of the external magnetic
field are shown. It can be seen that the critical beam current
$I_{SCL}$ decreases monotonically with $R/L$, and that for different
external magnetic fields the dependence $I_{SCL}(R)$ remains
essentially the same.}

{It should be noted that the optimal magnetic field $B_{\min}$, at
which the critical beam current is minimal, does not essentially
depend on the filling parameter $\sigma$ and geometry parameter
$R/L$ and is determined only by the velocity $\beta_0$. For
different values of the parameters $\sigma$ and $R/L$ the critical
beam current accepts minimal value for the optimal external magnetic
field $B = B_{\min}$ (see curves~{\sl2} in Figs.~\ref{fgr2}(b)
and~\ref{fgr2_2ADD}(b)).}

\section{Physical processes in the electron beam with overcritical
current at the external magnetic field}
\label{Section:PhysicalProcesses}

Let us consider the physical processes that accompany the
oscillating VC formation in the electron beam and govern the
dependence $I_{SCL}(B)$ described in previous
Section~\ref{Section:CriticalCurrentDependence} for different
inductions $B$ of the guiding magnetic field.

\begin{figure}
\centerline{\includegraphics*[scale=0.4]{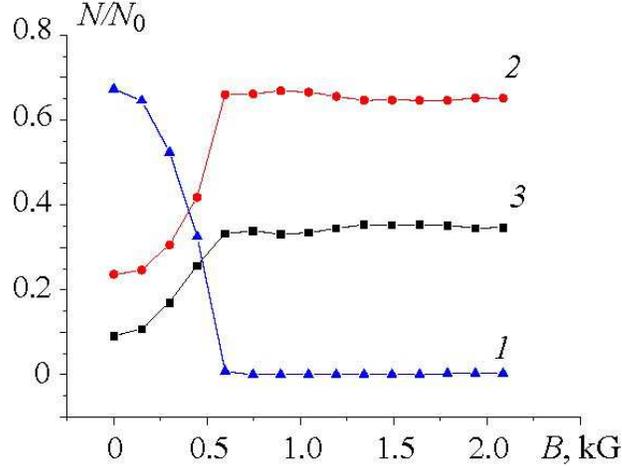}} \caption{(a)
Normalized number $N/N_0$ of charged particles escaping from the
drift space through the side wall (curve {\sl1}) and through the
entrance and exit grid electrodes (curves {\sl2} and {\sl3},
respecticely) versus external magnetic field for $\beta_0 = 0.5$ and
$\sigma = 0.5$. The value $N_0$ is the total number of injected
charged particles\label{fgr3}}
\end{figure}

We begin by analyzing the dependence of the beam current
distribution in the system on the external magnetic field. To do
this, we consider how the numbers of large charged particles that
escape from the drift space through its entrance and exit planes and
through its side surface depending on the external magnetic field.
Fig.~\ref{fgr3} shows how the normalized numbers $N$ of particles
that are reflected from the VC and escape from the drift space
through the entrance grid (curve {\sl2} in Fig.~\ref{fgr3}) and
through the side surface (curve {\sl1}) and that pass through the VC
and escape from the drift space through the exit grid (curve {\sl3})
depend on the axial magnetic field in the drift space. The
calculations were carried out for the filling parameter $\sigma =
0.5$ and for the beam velocity $\beta_0 = 0.5$.

From Fig.~\ref{fgr3} we can see that the escape dynamics of the
charged particles from the drift space differs radically in the
cases of the weak and the strong values of the external magnetic
field. In the case of the weak magnetic field $B < 0.5$\,kG, a
large number of charged particles escape from the drift space
through its side wall. The number of such particles, that are
reflected from the VC and escape from the system in the radial
(transverse) direction, is maximum at zero external magnetic field
and decreases monotonically with magnetic induction increasing
(see curve~{\sl1} in Fig.~\ref{fgr3}). At the same time, the
number of particles that are reflected from the VC in the
longitudinal direction and escape through the injection plane
$z=0$, as well as the number of particles that pass through the
oscillating VC and escape from the drift space through the exit
grid, increases with the external magnetic field (see
Fig.~\ref{fgr3}, curves {\sl2} and {\sl3} respectively). In
Fig.~\ref{fgr1} such behavior of the number of charged particles
that are reflected from the VC in the longitudinal and transverse
directions corresponds to the range $B<B_{\min}$ in which the
critical beam current decreases with increasing external magnetic
field (cf. the curves in the range $B < 0.5$\,kG in
Fig.~\ref{fgr1} versus the curves in the same range in
Fig.~\ref{fgr3}).

For stronger magnetic fields, $B > 0.5$\,kG there are almost no
charged particles that are reflected from the VC in the transverse
direction and escape radially from the drift space through its side
wall (see Fig.~\ref{fgr3}). The number of particles moving in the
longitudinal direction and escaping from the drift space through the
entrance and exit grids remains unchanged with the increasing of the
external magnetic field (see Fig.~\ref{fgr3}, curves {\sl2} and
{\sl3}). A comparison between Figs.~\ref{fgr1} and~\ref{fgr3} shows
that the magnetic field at which the transverse dynamics of the
charged particles reflected from the VC becomes unimportant (i.e.
$N\approx0$) and is close to the optimal magnetic field $B_{\min}$
at which the critical space charge limiting current is minimum.

Such behavior of the electron beam allows us to suggest that at the
weak external magnetic field the formation and dynamics of the VC
oscillations differ from those at the strong magnetic field. In the
weak magnetic field $B < B_{\min}$, which poorly confines the
electrons, the transverse dynamics of the electron beam plays a
predominant role in the drift space. For the strong external
magnetic field that the current through the side surface bounding
the drift tube in the radial direction is zero the longitudinal
dynamics of the space charge of the beam  is dominated in the
analyzed system. Let us note that cyclotron rotation of electrons in
the magnetic field is observed simultaneously. At the intermediate
magnetic fields $B\sim B_{\min}$ the behavior of the system is
governed by both the transverse and longitudinal dynamics of the
space charge in the VC region.

\begin{figure}
\centerline{\includegraphics*[scale=0.4]{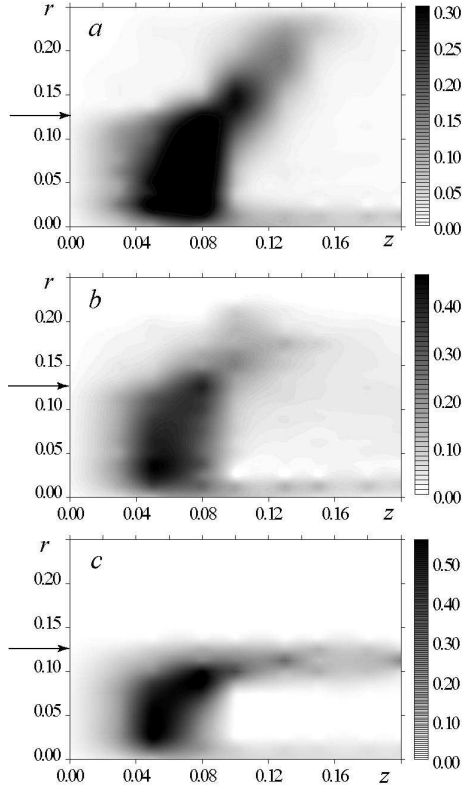}} \caption{Space
charge density distributions in the VC region, averaged over the
characteristic period of VC oscillations for the filling parameter
$\sigma = 0.5$ and different inductions of the external magnetic
field: (a) $B = 0$, (b) $B = B_{\min} = 0.6$\,kG, and (c) $B =
2.0$\,kG. The arrows show the initial radial beam boundary.
Longitudinal coordinate is shown on the abscissa and radial
coordinate is shown on the ordinate. The gray color intensity is
proportional to the absolute value of the space charge. The scale
from the right side of the figure shows the values of the space
charge density. \label{fgr4}}
\end{figure}

Let us examine in more detail the behavior of the electron beam in
the VC area at different inductions of the external magnetic
field. The space charge density distributions in the area of VC
oscillations ($z\in(0,0.2)$) averaged over the characteristic
period of VC oscillations for different inductions $B$ is shown in
Figs.~\ref{fgr4}.
In these figure the absolute values of the dimensionless space
charge density $|\rho(r,z)|$ at different points $(r, z)$ in the
drift space are shown by shades of gray. The maximum in the space
charge density (the dark region in the distributions $|\rho(r,
z)|$) corresponds to the VC region, where the electrons are
stopped and start moving back toward the injection plane or toward
the side wall of the drift tube. The electron beam is decelerated
by the space charge of the VC and the electron velocities within
it are close to zero. So, the space charge density $|\rho|$
achieves the maximum value in the VC region.

From the distributions shown in Figs.~\ref{fgr4}(a), which were
calculated for the zero external magnetic field, we can see that
in the VC region the electron beam greatly expands in the
transverse (radial) direction, so the beam space charge density
near the injection plane is nonzero over the entire cross
transverse section of the drift tube. For the weak external
magnetic field the space charge density distribution is
qualitatively the same. This behavior of the electron beam can be
explained as follows. A dense electron bunch formed in the VC
region (and, accordingly, decelerates the electron beam) gives
rise to fairly strong space charge forces (the Coulomb repulsion
between the electrons), under the effect of which the electrons
acquire the transverse velocity, rapidly escape in the radial
direction, and are lost at the side cylindrical wall bounding the
drift tube.

Such a transverse electron dynamics under the effect of space charge
forces in the absence of the external magnetic field or in the weak
external magnetic field, which does not restrain these forces,
results in substantial radial expansion of the beam. As a
consequence, in the weak external magnetic field most of the beam
electrons in the VC region move in the transverse (radial) direction
with respect to the drift tube axis. The radial beam expansion
reduces the space charge density in the VC and, accordingly,
increases the space charge limiting current $I_{SCL}$ for achieving
the space charge density required for the formation of the
oscillating VC. In this case some of the electrons begin to be
reflected to the entrance grid and are turned to the side surface of
the drift tube.

As can be seen from Fig.~\ref{fgr4}(a), the electron bunch in the
VC region extends over the entire transverse cross section of the
drift tube. In this case, the space charge density is maximum
($|\rho| > 0.25$ in dimensionless units) within the radial region
whose radius exceeds the initial beam radius $r_b$ (shown by the
arrow in Fig.~\ref{fgr4}(b)).  The radial region of maximum space
charge density is displaced in the propagation direction of the
injected electron beam. This indicates that the distribution
$\rho(r,z)$ in the VC region is nonuniform in both the radial and
longitudinal directions.

For the small filling parameter $\sigma$ the system behaves in the
similar manner. In this case the lower critical current of
electron beam is determined by the higher space charge density of
the electron beam because of the smaller beam cross-sectional area
for the fixed parameter $\alpha$ which is proportional to the
injected beam current $I$. Consequently, the space charge density
level required for the formation of the VC oscillations in the
beam of larger radius is achieved when the electron beam current
is higher than that of the beam with smaller radius. This
conclusion, which is confirmed by numerical simulations (see
Fig.~\ref{fgr2}), is valid for an arbitrary induction $B$ of the
external magnetic field.

Let us note, that for the weak magnetic field the electron bunch in
the VC region oscillates mainly in the radial direction. In this
case the maximum space charge density and the spatial position of
the VC vary only slightly with time, in contrast to the case of
strong external magnetic fields, in which the electron dynamics is
essentially one-dimensional (see, e.g.,
\cite{Sullivan:1987_VCO_Book, Birdsall:1966_book,
Anfinogentov:1998_VCO_ENGL}). Radial oscillations can occur with the
lower beam space charge densities than those at which the VC can
execute longitudinal oscillations.

A stronger magnetic field $B$ prevents the beam electrons moving
in the radial direction and imparts the azimuthal velocity to
them. As a result, the number of electrons escaping from the drift
space radially through the side wall of the system decreases as
the magnetic induction $B$ increases (see Fig.~\ref{fgr3}).
Fig.~\ref{fgr4}(b) shows the averaging space charge density
distributions in the VC region for the optimal magnetic field $B =
B_{\min}$. It is possible to see that the external magnetic field
restricts radial electron motion, increasing the space charge
density in the VC region (cf. the gray scale on the right of
Figs.~\ref{fgr4}(a) and~\ref{fgr4}(b)). Consequently, as follows
from Fig.~\ref{fgr1} decreasing the critical current at which the
electrons begin to be reflected and the oscillating VC arises in
the beam.

Fig.~\ref{fgr4}(c) shows that in the strong magnetic field such
that the beam dynamics close to one-dimensional ($B\gg B_{\min}$),
the VC (the region of maximum space charge density) ranges almost
entirely within the initial beam radius. The initial beam radius
$r_b$ is indicated by the arrow in Fig.~\ref{fgr4}. In this case
radial oscillations of the space charge density are not observed.
At strong magnetic fields the oscillatory behavior of the VC
agrees well with the VC oscillations in the one-dimensional case
described in detail in Refs.~\cite{Matsumoto:1996,
Jiang1995_Mech_VC, Koronovskii:2002_PierceEnglish} for the 1D
model of the electron beam with the VC. As consequence, VC
oscillations behavior and the space charge limiting current weakly
depends on the external magnetic field in the range $B \gg
B_{\min}$.

Based on the fact that the space charge dynamics differs radically
in the cases of the weak and the strong external magnetic field, we
may speak of two different types of oscillating VC. In the weak
magnetic field, the formation of the VC and its oscillations are
governed primarily by the transverse dynamics of the beam electrons
under the effect of the repulsive space-charge forces, which
substantially expands the beam in the radial direction. In this case
we may speak of the {\it transverse} VC in which the space charge
density oscillates predominantly in the radial direction. The
situation is radically different for the strong external magnetic
fields when the longitudinal electron dynamics is dominated and the
space charge density oscillates predominantly in the longitudinal
direction. In this case we may speak of the {\it longitudinal} VC in
an electron beam with an overcritical current. The behavior of such
beams with the longitudinal VC oscillations has been studied in more
detail by the numerical simulation \cite{Jiang:1996_VCO,
Jiang:1995_VircatorKlystron, Kwan:1988_reditron1,
Lin:1990_VC_electrostatic, Anfinogentov:1998_VCO_ENGL}. Such
oscillating regimes are typical of vircators operating with a strong
external magnetic field \cite{Lin:1990_VC_electrostatic,
Dubinov:2002_ReviewEngl}. Accordingly, the critical beam current is
described by the familiar relationship (\ref{I_0_B=INFTY})
\cite{Sullivan:1987_VCO_Book}.

At intermediate external magnetic fields $B\sim B_{\min}$
competition between the longitudinal and transverse oscillations of
the space charge density in the electron beam with the VC gives rise
to the minimum in the dependence $I_{SCL}(B)$ of the space charge
limit current on the external magnetic field.

\section{Conclusion}
\label{Section:Conclusion}

Thus, the influence of the external magnetic field on the space
charge limiting current corresponding to the onset of the
oscillating VC in the electron beam is reported for the first time
in this Letter. It is shown that the critical current decreases when
external magnetic field increases. There is the optimal magnetic
field at which the space charge limiting current is minimum. For
stronger external magnetic fields the critical current is described
by the familiar relation (\ref{I_0_B=INFTY})
\cite{Sullivan:1987_VCO_Book} obtained under the assumption that the
motion of the electrons is one-dimensional. Study of the physical
processes in the electron beam with the virtual cathode presents
that the dependence of the critical current on the external magnetic
field is governed by the characteristic features of the longitudinal
and radial dynamics of the electron space charge in the drift tube
at different inductions of the axial magnetic field.

In the weak external magnetic field the major role is played by the
transverse beam dynamics (i.e. the two-dimensional effects
associated with the Coulomb repulsion of the electrons in the radial
direction). As a consequence, the space charge density in the VC
area reduces, so the oscillating VC can arise in the electron beam
with the higher current in comparison with that at strong magnetic
fields. When the external magnetic field is strong the transverse
beam dynamics is less important and the behavior of the VC is
governed primarily by the longitudinal oscillations in the electron
beam.

\section*{Acknowledgments}
\label{Sct:Acknowledgments}

We thank Dr. S.V. Eremina for English language support. This work
was supported by U.S. Civilian Research and Development Foundation
for the Independent States of the Former Soviet Union (CRDF),
grant REC--006, the Supporting program of leading Russian
scientific schools (project NSh--4167.2006.2) and Doctor of
Science (project MD-1884.2007.2), Russian Foundation of Basic
Research (05-02-16286 and 06-02-81013) and ``Dynasty'' Foundation.




\bibliographystyle{elsart-num}

\end{document}